\documentclass[11pt,twoside]{article}
\usepackage{./asp2010}

\resetcounters

\aspvolume{TBD}
\aspcpryear{2013}
\aspvoltitle{Structure and Dynamics of Disk Galaxies}
\aspvolauthor{Marc S.~Seigar and Patrick Treuthardt, eds.}

\def\Msun{\> M_{\odot}}
\def\kpc{\> {\rm kpc}}
\def\Gyr{\> {\rm Gyr}}

\begin{document}

\title{Local Group and Star Cluster Dynamics from HSTPROMO\\ 
(The Hubble Space Telescope Proper Motion Collaboration)}


\author{%
Roeland P.~van der Marel,$^1$ 
Jay Anderson,$^1$
Andrea Bellini,$^1$
Gurtina Besla,$^2$
Paolo Bianchini,$^3$
Mike Boylan-Kolchin,$^4$
Julio Chaname,$^5$
Alis Deason,$^6$
Tuan Do,$^7$
Puragra Guhathakurta,$^6$
Nitya Kallivayalil,$^8$
Danny Lennon,$^9$
Davide Massari,$^{10}$
Eileen Meyer,$^1$
Imants Platais,$^{11}$
Elena Sabbi,$^{1,9}$
Sangmo Tony Sohn,$^1$
Mario Soto,$^1$
Michele Trenti,$^{12}$
Laura Watkins$^1$
%
%
\affil{$^1$Space Telescope Science Inst., 3700 San Martin Drive,
Baltimore, MD 21218}
\affil{$^2$Dept.~of Astronomy, Columbia University, New York, NY 10027}
\affil{$^3$Max Planck Inst. for Astronomy, K\"{o}nigstuhl 17, 69117 
Heidelberg, Germany}
\affil{$^4$Dept.~of Astronomy, University of Maryland, College Park, 
MD 20742}
\affil{$^5$Departamento de Astronomia y Astrofisica, Pontificia Universidad 
Catolica de Chile, Av. Vicuna Mackenna 4860, 782-0436 Macul, Santiago, Chile}
\affil{$^6$Dept. Astronomy \& Astrophysics, Univ.~of California,
Santa Cruz, CA 95064}
\affil{$^7$Dunlap Institute for Astronomy and Astrophysics, University
of Toronto, Toronto, Ontario, Canada M5S 3H4}
\affil{$^8$Dept.~of Astronomy, University of Virginia, Charlottesville, 
VA 22904}
\affil{$^9$ESA-ESAC, Apdo. de Correo 78, 28691 Villanueva de la Canada, Spain}
\affil{$^{10}$Dipartimento di Fisica e Astronomia, Universita degli
Studi di Bologna, Viale Berti Pichat 6/2, I-40127 Bologna, Italy}
\affil{$^{11}$Department of Physics \& Astronomy, Johns Hopkins
University, 3400 N.~Charles Street, Baltimore, MD 21218}
\affil{$^{12}$Kavli Institute for Cosmology, University of Cambridge,
Madingley Road, Cambridge, CB3 0HA, UK}
}


\begin{abstract}
The Hubble Space Telescope (HST) has proven to be uniquely suited for
the measurement of proper motions (PMs) of stars and galaxies in the
nearby Universe. Here we summarize the main results and ongoing
studies of the HSTPROMO collaboration, which over the past decade has
executed some two dozen observational and theoretical HST projects on
this topic. This is continuing to revolutionize our dynamical
understanding of many objects, including: globular clusters; young
star clusters; stars and stellar streams in the Milky Way halo; Local
Group galaxies, including dwarf satellite galaxies, the Magellanic
Clouds, and the Andromeda galaxy; and AGN Black Hole Jets.
\end{abstract}



\section{Introduction: Proper Motion Studies with HST}

The dynamics of stars, clusters, and galaxies provide important
information on the formation, evolution, structure, and mass of
stellar systems. Most of what is known is based on observations of
line-of-sight (LOS) velocities. Such observations constrain only one
component of motion, and interpretation therefore generally requires
that various assumptions be made. To determine fully three-dimensional
velocities, it is necessary to also determine PMs. If a PM accuracy
$\Delta$ PM $\approx 50 \> \mu$as/yr is achieved, then many dynamical
topics in the Local Group can be meaningfully addressed. This
corresponds to a velocity accuracy $\Delta v \approx (D/4) $ km/s at
distance $D$ kpc.

Such PM accuracies are not generally accessible from the ground. VLBA
can reach them, but only for a small number of maser sources. Gaia
should reach them for many stars, but this will take several more
years. For HST, $50 \> \mu$as/yr corresponds to a motion of $\sim
0.01$ CCD pixel over 10 years. Since HST has many advantages for
astrometry (high spatial resolution, long-term stability, and more
than 20 years of Archival data), such accuracies have already been
routinely achievable for years.  This is true even for faint sources
in crowded fields (unlike Gaia), thus allowing PM measurements for $N
= 10^2$--$10^6$ sources per field, depending on the specific target.


The random errors for measurements of bulk motions and velocity
dispersions scale as $N^{-0.5}$. Thus, systematic errors are often the
limiting factor. These can be controlled by careful calibrations of PSF
shapes, geometric distortions, charge-transfer efficiency, color
effects, etc.~\citep{2000PASP..112.1360A}. Relative PM measurements
suffice for measurements of velocity dispersions, while absolute PM
measurements are required for bulk motion measurements. The latter
requires use of distant (i.e., stationary) background quasars or
galaxies as reference sources \citep{2008PASP..120..907M}.\looseness=-2


Several other groups have used HST for PM studies of globular clusters
\citep[e.g.,][]{2003ApJ...595..187M,2012ApJ...745..175M,2006ApJS..166..249M},
the Milky Way
\citep[e.g.,][]{2002AJ....124.2054K,2010ApJ...719L..23B}, or the Local
Group \citep[e.g.,][]{2008IAUS..248..244P,2011ApJ...741..100L}.  Here
instead we report on results and ongoing work of our HSTPROMO
collaboration, the work of which has included some two dozen
observational and theoretical HST projects over the past
decade.\footnote{See details on the HSTPROMO web page:
  http://www.stsci.edu/$\sim$marel/hstpromo.html}

\section{HSTPROMO Results and Projects}

The following provides a sampling of our results on PM dynamics in the
nearby Universe, roughly in order of increasing distance.


\vspace{-0.4truecm}
\paragraph{Globular Clusters (GCs)}  

\citet{2010ApJ...710.1032A} presented a PM catalog for 170,000 stars
in the GC omega Centauri. Detailed dynamical models by
\citet{2010ApJ...710.1063V} placed an upper limit of $1.2 \times 10^4
\Msun$ on the mass of any possible intermediate mass black hole (IMBH)
at the center of this GC. Comparison of the PM dispersion of stars of
different mass along the main sequence (MS) shows that Omega Cen is
not in equipartition. \citet{2013arXiv1302.2152T} performed $N$-body
simulations to model this, and found that despite popular belief to
the contrary, GCs are not generally expected to ever get close to
equipartition. \citet{Massari13} presented measurements of the
absolute PM of the GC NGC 6681 (M70). \citet{2013MmSAI..84..140B} are
extending these studies to a sample of two dozen GCs. The resulting
datasets can be interpreted with dynamical modeling tools such as
those presented by \citet{2008ApJ...682..841C}, and will provide many
new insights into the structure, dynamics, multiple populations and
possible IMBHs of GCs.


\vspace{-0.4truecm}
\paragraph{Young Star Clusters} 

The phenomenon of runaway O/B stars from young clusters has been known
for a long time. To better understand the dynamical origin of these
stars, we are determining PMs of stars in the 30 Dor star-forming
region in the Large Magellanic Cloud \citep[LMC; ][Platais et al., in
  prep.]{2013AJ....146...53S}, and in regions near young clusters in
the Galactic Center (Lennon et al., in prep.). An important advantage
of such measurements is that the direction of the PM vectors can help
identify the probable places of origin of any runaway stars.

 
\vspace{-0.4truecm}
\paragraph{Milky Way (MW) Halo Stars and Streams} By combining PM and 
color-magnitude diagram information it is possible to uniquely
identify distant MS halo stars. The PMs of 13 stars at $24 \pm 6 \kpc$
toward the Andromeda-Triangulum halo overdensity imply a halo velocity
ellipsoid that is more tangentially anisotropic than near the
Sun. This suggests the presence of a shell in the halo, probably
resulting from an ancient accretion event \citep{2013ApJ...766...24D}.
We will extend this work to a sample of $\sim 1000$ stars spread over
the sky. We are also in the process of determining the PMs for stars
along the Sagittarius and Orphan Streams (Sohn et al., van der Marel
et al., in prep.), which will constrain the shape of the MW dark
matter halo.


\vspace{-0.4truecm}
\paragraph{Magellanic Clouds}

\citet{2006ApJ...652.1213K,2006ApJ...638..772K} were the first to
obtain high-accuracy PMs of the Magellanic Clouds, which were recently
refined by \citet{2013ApJ...764..161K}. The results imply that the
Clouds are moving faster than previously believed, and are likely just
past their {\it first} MW pericenter \citep{2007ApJ...668..949B}. This
has led to revisions in our understanding of both the Magellanic
Stream \citep{2010ApJ...721L..97B}, and the formation of Magellanic
Irregular galaxies in general \citep{2012MNRAS.421.2109B}. By mapping
the variations in PM across the face of the LMC, it has been possible
to measure its PM rotation field and rotation curve
\citep{2013arXiv1305.4641V}, the first time this has been possible for
any galaxy.



\vspace{-0.4truecm}
\paragraph{Local Group (LG) Dwarf Satellite Galaxies}

\citet{2013ApJ...768..139S} measured the absolute PM of the
rapidly-moving distant MW satellite Leo~I, which indicates it likely
had its {\it first} pericenter passage $1.0 \pm 0.1 \Gyr$ ago at $91
\pm 36 \kpc$. Cosmological simulations show that subhalos are almost
always bound to their host. Assuming that Leo~I is the least-bound
classical MW satellite, the implied MW virial mass is $M_{\rm vir}
\approx (1.6 \pm 0.3) \times 10^{12} \Msun$
\citep{2013ApJ...768..140B}. We are also studying other LG dwarfs, to
determine both internal PM kinematics (Sohn et al., in prep.: Draco,
Sculptor) and absolute PMs \citep[][for the Sgr dSph; van der Marel et al.,
  in prep: Cetus, Tucana, Leo A, Sgr dIrr; Do et al., in prep.:
  Leo~T.]{Massari13}

\vspace{-0.4truecm}
\paragraph{The Andromeda Galaxy (M31)}

\citet{2012ApJ...753....7S} obtained the first-ever PM measurement for
M31. The result agrees with indirect estimates based on the LOS
kinematics of its satellites \citep{2008ApJ...678..187V}. The LG
timing argument combined with other mass estimates implies a total LG
virial mass $M_{\rm MW} + M_{\rm M31} = (3.2 \pm 0.6) \times 10^{12}
\Msun$ \citep{2012ApJ...753....8V}. The PM is consistent with a
head-on collision orbit for M31 toward the MW. $N$-body simulations
show that the first passage will occur in 4 Gyr, followed by a
complete merger after 6 Gyr \citep{2012ApJ...753....9V}. The remnant
will resemble an elliptical galaxy. There is a 10\% probability that
the Triangulum galaxy, M33, will hit the MW before M31
does.\looseness=-2

\vspace{-0.4truecm}
\paragraph{AGN Black Hole Jets}

\citet{2013ApJ...774L..21M} measured the PMs of features in the
optical jet of M87, and found evidence for helical motion. Similar
measurements for 3C264, 3C273, and 3C346 are in progress.

\vspace{0.2truecm}
\acknowledgements Support for individual HSTPROMO projects is provided
by NASA through grants from STScI, which is operated by AURA, Inc.,
under NASA contract NAS 5-26555.

\vspace{-0.2truecm}
\bibliography{author}

\end{document}